\documentclass[aip,twocolumn,preprintnumbers,superscriptaddress,showpacs]{revtex4}
\usepackage{amssymb}
\usepackage{graphicx}
\usepackage{dcolumn}
\usepackage{verbatim}
\usepackage{bm}
\usepackage{multirow}
\usepackage{color}

\usepackage{epstopdf}

\begin{document}
\title{Room temperature multiferroism in CaTcO$_3$ by interface engineering}

\author{Hongwei Wang}
\affiliation{Key Laboratory of Quantum Information, University of Science and Technology of 
China, Hefei, Anhui, 230026, People's Republic of China}
\affiliation{Department of Physics,
Temple University, Philadelphia, PA 19122, USA}
\author{Lixin He$^*$}
\affiliation{Key Laboratory of Quantum Information,
University of Science and Technology of China, Hefei, 
Anhui, 230026, People's Republic of China}
\author{Xifan Wu$^*$}
\affiliation{Department of Physics, 
Temple University, Philadelphia, PA 19122, USA}
\date{\today }

\begin{abstract}

The structural instabilities of ATcO$_3$(A= Ca, Sr, Ba) are investigated by first-principles calculations. 
In addiction to the large octahedral rotation instability, a weak ferroelectric 
tendency is identified in CaTcO$_3$. We show that the ferroelectricity in CaTcO$_3$
can be recovered by interface engineering based on CaTcO$_3$/BaTcO$_3$ superlattices, where
the octahedral rotation is largely suppressed. The N{\'e}el temperature 
is found to be $\sim$ 816 K, indicating that CaTcO$_3$ can be 
engineered into a new room temperature multiferroic material. 

\end{abstract}

\maketitle

ABO$_3$ perovskite continues to prove itself to be an important family of multifunctional materials.
This is because various instabilities can be simultaneously present within its simple cubic
structure at high temperature~\cite{Hwang_NM_2012, Jiawang_PRB_2013, Joe_PRL_2013,
Junhee_PRB_2011,Junhee_PRL_2011}. 
Many of these instabilities such as magnetic, ferroelectric, antiferrodistortive,
and antiferroelectric orderings and their interactions are closely associated with different functionalities
that are useful for the device applications. Recently 
a lot of excitement has been generated, again, in ATcO$_3$(A= Ca, Sr, Ba) 
as a new family of perovskite~\cite{CaTcO3_JACS_2011,RTcO3_PRB_2011, SrTcO3_PRL_2011}.
Similar to its isovalent neighbor of Mn($4s^23d^5$), technetium has a $5s^24d^5$ electronic configuration
with a completely occupied $t_{2g}$ and completely empty $e_g$ bands according to the crystal field theory
under cubic symmetry. In the crystalline phase, ATcO$_3$ is paraelectric and adopts a G-type
antiferromagnetic (AFM) configuration as expected. Strikingly, ATcO$_3$ is discovered to
have an anomalously high magnetic ordering temperature,
e.g. $T_N \sim 800$ K in CaTcO$_3$~\cite{CaTcO3_JACS_2011,RTcO3_PRB_2011, SrTcO3_PRL_2011}. 
This is in sharp contrast to that of manganese perovskite, in which
CaMnO$_3$ only has a Ne{\'e}l temperature of 123 K~\cite{CaMnO3_Curie, CaMnO3_PRB_2012}. 
Keeping this intriguing property in mind,
one might be wondering whether more functionalities could be explored in this new family of perovskite?

In this work, we first use density functional theory (DFT) to perform an investigation of
the structural instabilities in ATcO$_3$.
An unexpected hidden ferroelectric (FE) instability is discovered in CaTcO$_3$, along
with a much stronger antiferrodistortive (AFD) one in its cubic phase at high temperature. 
Based on a recently introduced interface engineering mechanism~\cite{Hongwei_EPL},
we carry out DFT calculations in CaTcO$_3$/BaTcO$_3$ supercells and explicitly show that the FE polarization
can be recovered in interfacial layer, where the AFD instability 
associated with oxygen octahedral rotation  
is largely suppressed.
The Ne{\'e}l temperature of the supercell is identified to be $\sim$ 816 K by a
following Monte Carlo simulation. 
Our theoretical predictions clearly suggest that the 
room temperature multiferroism can be achieved in CaTcO$_3$ by interface engineering.

Our DFT calculations are done by using the VASP code package~\cite{kresse93,kresse96}.
In particular, both hybrid DFT within the Heyd-Scuseria-Ernzerhof (HSE) scheme~\cite{HSE}
and the Perdew-Burke-Ernzerhof functional
revised for solids (PBEsol)~\cite{perdew08} functional are adopted to 
approximately treat the exchange correlation of electrons~\cite{HSE_PBE_U}.
These DFT functionals are known to be very accurate in predicting
the volume of solids which is particularly important because the FE instability
is sensitive to the cell volume. 
The lattice constants are fully relaxed in bulk
studies and allowed to relaxed in the [001] direction in CaTcO$_3$/BaTcO$_3$ supercells,
assuming an epitaxial growth on the substrate of GdScO$_3$.

Although the detailed experimental study of the structural phase transition 
is only available for SrTcO$_3$~\cite{SrTcO3_PRL_2011} and CaTcO$_3$~\cite{CaTcO3_JACS_2011}.
It is generally accepted that ATcO$_3$ share a common orthorhombic Pnma
symmetry { at low temperature} and cubic
${\rm Pm\bar{3}m}$ symmetry at { high temperature}.
In Table I, we present the computed structural parameters, in which a direct comparison
is also made with experiments of the low temperature phase.
Overall, it can be seen that PBEsol functional gives a very accurately prediction of
lattice constants and volume, except that the lattice constant $a$ is slightly
overestimated by $\sim$ 0.7\%.
The HSE functional gives even better agreement with the experimental values.
In the low temperature orthorhombic structure, the TcO$_6$ octahedral is allowed to both rotate and
tilt. It is interesting to note that with the increasing radius of $A$ site
atom, the rotational and tilting angles are decreasing
from very large angle in CaTcO$_3$ to almost zero in BaTcO$_3$.
This trend is very similar to that of ATiO$_3$, which
can be explained in terms of the so-called Goldschmidt tolerance factor~\cite{Goldschmidt}, in which a
small(large) tolerance factor favors(resists) octahedral rotation.

\begin{table}[ht]

\caption{Computed structural parameters, magnetic exchange angles$(^{\circ})$, oxygen octahedral 
rotation $\phi_{\rm r}(^{\circ})$ and tilting angles $\phi_{\rm t}(^{\circ})$
of bulk ATcO$_3$(A=Ca,Sr,Ba) with $Pnma$ symmetry. The structural parameter, Born effective charges $Z^*$, and electronic 
contribution to dielectric constant $\epsilon_\infty$ are also reported for $\rm {Pm\bar{3}m}$ symmetry. 
Experimental values~\cite{RTcO3_PRB_2011} are shown in parenthesis.}
\label{tab:calculations}\centering
\begin{tabular}{c|c|c|c|c}
\hline\hline
&                        &\scriptsize CaTcO$_3$ &\scriptsize SrTcO$_3$ &\scriptsize BaTcO$_3$ \\ \hline
\multirow{6}{10.5mm}{Pnma \\ \scriptsize (HSE)} &\scriptsize $a(\AA)$ & \footnotesize5.53 (5.53)    &\footnotesize5.56  (5.54)  &\footnotesize 5.69  \\
                         &\scriptsize $b(\AA)$ &\footnotesize 7.71 (7.70)    &\footnotesize7.86  (7.85)  &\footnotesize 8.05  \\
                         &\scriptsize $c(\AA)$ &\footnotesize 5.39 (5.39)    &\footnotesize5.61  (5.58)  &\footnotesize 5.67  \\
                         &\scriptsize $V(\AA^3)$ &\footnotesize 230.05(229.15)   &\footnotesize244.92 (242.74) &\footnotesize 260.20 \\
                         &   \tiny${\rm Tc}-\widehat{\rm O_{1}}-{\rm Tc}$ &\footnotesize150.90(150.43)   &\footnotesize163.39 (161.57) &\footnotesize 179.96 \\
                         &  \tiny ${\rm Tc}-\widehat{\rm O_{2}}-{\rm Tc}$ &\footnotesize150.91(151.53)   &\footnotesize167.76 (166.96) &\footnotesize 179.90 \\ 

                         &\scriptsize  $\phi_{\rm r}$ &\footnotesize 10.04   &\footnotesize 0.60  &\footnotesize 0.049 \\
                         &\scriptsize  $\phi_{\rm t}$     &\footnotesize 14.55   &\footnotesize 8.32  &\footnotesize 0.019 \\ \hline
                         
\multirow{6}{10.5mm}{Pnma \\ \scriptsize (PBEsol \\ +U)}   &\scriptsize $a(\AA)$ &\footnotesize 5.57 (5.53)    &\footnotesize 5.59 (5.54) & \footnotesize 5.71  \\
                             &\scriptsize $b(\AA)$ &\footnotesize 7.73 (7.70)    &\footnotesize7.90  (7.85)  &\footnotesize 8.08  \\
                             &\scriptsize $c(\AA)$ &\footnotesize 5.41 (5.39)    &\footnotesize5.60  (5.58)  &\footnotesize 5.71  \\
                             & \scriptsize $V(\AA^3)$&\footnotesize 232.08(229.15)   & \footnotesize 247.51 (242.74) & \footnotesize 262.87 \\
& \tiny${\rm Tc}-\widehat{\rm O_{1}}-{\rm Tc}$  &\footnotesize148.41(150.43)   &\footnotesize161.13 (161.57) &\footnotesize 179.94 \\
&  \tiny${\rm Tc}-\widehat{\rm O_{2}}-{\rm Tc}$ &\footnotesize148.91(151.53)   &\footnotesize161.24 (166.96) &\footnotesize 179.88 \\ 
                         &\scriptsize  $\phi_{\rm r}$ &\footnotesize 10.63   &\footnotesize 6.18  &\footnotesize 0.058 \\
                         &\scriptsize  $\phi_{\rm t}$     &\footnotesize 15.80   &\footnotesize 9.13  &\footnotesize 0.032 \\ \hline
\multirow{6}{10.5mm}{$\rm {Pm\bar{3}m}$\\ \scriptsize (PBEsol \\ +U)} &\scriptsize $a(\AA)$ &\footnotesize 3.93  &\footnotesize 3.97 &\footnotesize 4.03 \\
                              & \scriptsize$Z^*_{\rm A}$ &\footnotesize 2.69  &\footnotesize 2.66 &\footnotesize 2.87 \\
                              & \scriptsize$Z^*_{\rm Tc}$  &\footnotesize 5.87  &\footnotesize 5.96 &\footnotesize 6.19  \\
                              & \scriptsize$Z^*_{{\rm O} \parallel}$ &\footnotesize $-$1.62 &\footnotesize $-$1.62 &\footnotesize $-$1.72 \\
                              & \scriptsize$Z^*_{{\rm O} \perp}$ &\footnotesize $-$5.32 &\footnotesize $-$5.38 &\footnotesize $-$5.61  \\
                              & \scriptsize$\epsilon_\infty$ &\footnotesize 8.64 &\footnotesize 8.60 &\footnotesize 9.14   \\ \hline\hline
\end{tabular}
\end{table}

The oxygen rotation in both CaTcO$_3$ and SrTcO$_3$ 
indicates that there is a structural instability located at Brillouin zone boundary
in their high temperature cubic phase.
However, it will be also interesting to check whether there is any additional instabilities that
might lead to functional properties. To this end, we
perform the phonon band structure calculations along high
symmetry points in the cubic phase of ATcO$_3$ by HSE hybrid functional.
As expected, we find that there is no unstable phonon in BaTcO$_3$,
and an unstable AFD phonon at $R$ point of Brillouin zone boundary
in SrTcO$_3$ ($\omega= 150i ~{\rm cm}^{-1}$),
which becomes much more unstable in CaTcO$_3$ ($\omega= 232.67i ~{\rm cm}^{-1}$).
Surprisingly, we also identify a hidden FE phonon mode in CaTcO$_3$ at zone
center ($\omega= 58.82i ~{\rm cm}^{-1}$),
however, weaker than that of { the AFD mode} as shown in Fig.~1.
This suggests that cubic CaTcO$_3$ has a tendency to develop nonzero
spontaneous polarization. 
In order to elucidate the nature of FE, we further compute the
Born effective charge~\cite{effective_charge} and the results are shown in Table.~1.
One can clearly see anomalously large effective charges on Tc and O$_{\parallel}$,
which are similarly identified in conventional ferroelectrics such as BaTiO$_3$.
This suggest that the covalent bonding between Tc and oxygen atoms lies at
the origin of the weak FE in CaTcO$_3$, in which a large dynamic charge transfer
results in a polar instability for the atomic displacement along the Tc-O bond direction~\cite{covalent}.
Unfortunately, the ferroelectric tendency does not result in a condensation of FE ordering
in its low temperature orthorhombic phase below 800 K, where CaTcO$_3$ stays paraelectric.
This is due to the competition between the FE and AFD instabilities~\cite{competition}. As a result,
the development of ferroelectric polarization is prohibited in the presence of a
strong tendency of oxygen octahedral rotation.
With the intriguingly high Ne{\' e}l temperature, it is particularly desirable that
the FE ordering could be recovered and turned into a new room temperature multiferroic material.
However, additional material engineering is required.


\begin{figure}[t]
\centerline{
\includegraphics[width = 3.0 in]{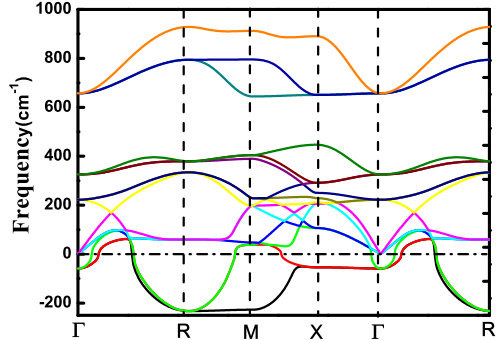}}
\vspace{-0.5cm}
\caption{ (Color online)
Phonon dispersions of the G-type AFM CaTcO3 of $\rm {Pm\bar{3}m}$ symmetry along the high-symmetry points
, which are computed by HSE functional.}
 \label{fig1}
\end{figure}

The above can be achieved in the artificial materials of superlattices under
epitaxial growth. In general, there are two kinds of approaches 
that can be used which are the {\it epitaxial strain} and {\it interface
engineering} methods respectively. We first try the epitaxial method. To do so, we apply both tensile and
compressive strain on the cubic CaTcO$_3$ systematically in the range of $\pm\sim5\%$.
At the same time, we lower the space group symmetry { in the calculations}
to allow the development of polarization along both $c$ axis and in the $ab$ plane.
For the epitaxial strain, the FE ordering usually has a much larger dependence on the
strain through the strain polarization coupling than that of AFD ordering associated
with oxygen rotation. It is expected that at certain strain, the
ordering of AFD and FE can be switched. However our calculations show that
CaTcO$_3$ persists to be paraelectric even as much as $\pm5\%$ has been applied.

We then focus on the possibility of interface engineering as the second option.
Compared with epitaxial strain, more dramatic change in the strength of
instabilities can be obtained at the interface resulting in an overall change
of functionalities. In particular, it has been recently shown that a substantial
reducing AFD mode  at the interface layer of CaTiO$_3$/BaTiO$_3$  can be
used to enhance the FE~\cite{Xifan_PRB_2011}. Thus it will be very interesting to check if FE can
be recovered in the CaTcO$_3$/BaTcO$_3$ short-period superlattices.
We carry out first-principles calculations for BaTcO$_3$/CaTcO$_3$ superlattices
assuming the coherent growth on the substrate of GdScO$_3$ ($a=3.97\AA$)~\cite{Substrate}.
This selection of substrate only introduces a small lattice mismatch by about
$\pm 1.26\%$ in BaTcO$_3$ and CaTcO$_3$, which facilitates the experimental
growth. The main results are shown in Table.~II.
As one of the most exciting results, the 1BaTcO$_3$/1CaTcO$_3$ superlattice does
exhibit a FE ordering with a spontaneous polarization $P_s= 6.34 ~\mu c/cm^2$ along the [110]
in-plane axis!

The emergence of FE ordering only in superlattice suggests that it is an interface
effect. In order to further elucidate its interfacial origin, we employ the layer
polarization decomposition to analyze the local polarization profile~\cite{effective_charge}.
The results are shown in  Table.~II  along with the local oxygen octahedral rotation profile.
Compared with the bulk CaTcO$_3$, it can be seen that both oxygen rotation
and tilting angles associated with AFD are greatly reduced at the interfaces.
As a result of the reduced AFD mode, we see the development of local polar distortion
at the interfaces. Very interestingly, the local polar distortion is developed both along
[110] and [001] directions. However, in the [001] direction the local polarization
adopts an antiferroelectric-like profile and the resulting total polarization is
almost zero.
The interface nature of the induced FE can be more easily seen in the local polarization
profile in 2BaTcO$_3$/2CaTcO$_3$ supercell. This is because both interface,
bulk BaTcO$_3$, and bulk CaTcO$_3$ layers can be found in 2:2 supercell. 
As a result, the local polarization of interface layers can be directly compared to
those of bulks. The resulting local polarization profile is also presented in Table.~II.
Clearly, it can be seen that the polar distortions are only developed at the interfaces
along both [110] and [001] directions. This is due to the interface suppression of AFD 
similar to what has been already seen in 1:1 supercell. However, as soon as 
the bulk BaTcO$_3$ or CaTcO$_3$ layer is reached, the local polarization disappears 
immediately. At the same time both oxygen octahedral rotation and tilting 
angles recover to the bulk magnitudes. This is consistent with the complete absence of FE instability in
BaTcO$_3$ and the dominant AFD instability over FE in CaTcO$_3$. We want to
stress that the FE will be only induced in the CaTcO$_3$ layer when the oxygen
rotation is suppressed which lies at the heart of our interface design.

\begin{table}[ht]
\caption{Layer-by-layer decompositions of polarization $p$($\mu c/cm^2$), 
oxygen octahedral rotation $\phi_{\rm r}(^{\circ})$ , and 
tilting $\phi_{\rm t}(^{\circ})$ in the supercells of 
1BaTcO$_{3}$/1CaTcO$_{3}$(1BT1CT) and 2BaTcO$_{3}$/2CaTcO$_{3}$(2BT2CT),
which are computed by PBEsol+U functional.}
\label{tab:2}\centering
\begin{tabular}{c|c|cccc}
  \hline
  \hline
  
   &         & {\footnotesize CT} & {\footnotesize Interface} & {\footnotesize BT} & {\footnotesize Interface} \\
  \hline
  \multirow{4}{12mm}{\scriptsize 1BT1CT  }
  &\scriptsize $p_{\rm [110]}$   & -  & \footnotesize 6.31 & - &\footnotesize 6.18    \\
  &\scriptsize $p_{\rm [001]}$    & - & \footnotesize 1.53 & -  &\footnotesize $-$1.59   \\
  &\scriptsize $\phi_{\rm r}$ & - &\footnotesize 5.82 & - &\footnotesize 5.94     \\
  &\scriptsize $\phi_{\rm t}$    & - &\footnotesize 12.76 & - &\footnotesize 6.05      \\
  \hline
  \multirow{4}{12mm}{\scriptsize 2BT2CT }
  &\scriptsize $p_{\rm [110]}$     &\footnotesize 0.15  &\footnotesize 9.42 &\footnotesize $-$0.072 &\footnotesize $-$9.06 \\
  &\scriptsize $p_{\rm [001]}$     &\footnotesize 0.045  &\footnotesize 1.56 &\footnotesize 0.058  &\footnotesize $-$1.56 \\
  &\scriptsize $\phi_{\rm r}$ &\footnotesize 9.76  &\footnotesize 5.97 &\footnotesize 0.46  &\footnotesize 5.77  \\
  &\scriptsize $\phi_{\rm t}$     &\footnotesize 14.63 &\footnotesize 9.52 &\footnotesize 4.40  &\footnotesize 9.50  \\
  \hline
  \hline
\end{tabular}
\end{table}

Both BaTcO$_3$ and CaTcO$_3$ have been found to have high  Ne{\'e}l temperature
well above room temperature. It has been argued that the more delocalized $4d$ orbital
enhance the covalent hybridization with neighboring oxygen atoms, which in turn
increase the hopping matrix according to the Anderson-Goodenough-Kanamori rules~\cite{CaTcO3_JACS_2011, Goodenough}.
Therefore, it can be expected that the large magnetic antiferromagnetic coupling
will be kept in the BaTcO$_3$/CaTcO$_3$ superlattices. To further confirm this,
we map the total energies in different magnetic configurations onto the Heisenburg
Hamiltonian~\cite{Junhee_PRB_2011}. The exchange coupling constants include the nearest-neighbor ($J^1$) and next
-nearest-neighbor ($J^2$), beyond which the coupling constant is found to be very small and could
be neglected. This coupling constants are further decomposed into the contributions
from intralayer ($J_{\rm intra}$) and interlayers ($J_{\rm inter}$). 
The results are listed in Table~III. One can see that the
exchange coupling constants are approximately the averages of those of the bulk BaTcO$_3$ and bulk CaTcO$_3$
in both intra- and inter- layer contributions. We obtain the AFM phase
transition at 816 K from the Monte Carlo simulation.

\begin{table}[ht]
\caption{Magnetic moments $m$($\mu_B$), exchange coupling constants $J$(meV), and Ne{\'e}l temperature T$_{\rm N}$(K) of 
bulk CaTcO$_3$, bulk BaTcO$_3$, and 1CaTcO$_3$/1BaTcO$_3$ supercell. PBEsol+U functional is used in the DFT calculations.}
\label{tab:2}\centering
\begin{tabular}{c|c|c|c}
\hline\hline
               & CaTcO$_3$ & BaTcO$_3$ & 1CaTcO$_3$/1BaTcO$_3$ \\ \hline
             $m$      &   2.14 &  2.09  & 2.05  \\
                           $J^1_{\rm inter}$   & $-$38.10 & $-$69.02 &$-$52.60 \\
                           $J^1_{\rm intra}$   & $-$38.10 & $-$69.02 &$-$60.29 \\
                           $J^2_{\rm inter}$   &  $-$0.78 & $-$1.09  &$-$2.89  \\
                           $J^2_{\rm intra}$   &  $-$0.78 & $-$1.09  &$-$3.17  \\
                          $T_{\rm N}$         &  602   &  1089  & 816   \\  \hline\hline
\end{tabular}
\end{table}

In conclusion, a hidden ferroelectric instability is identified in CaTcO$_3$, however
suppressed by the large oxygen octahedral rotation. We use the first-principles
calculations to show that the polarization can be induced by an interface engineering
method in the proposed BaTcO$_3$/CaTcO$_3$ superlattices. Our computational results
confirm the room temperature multiferroism in the superlattice.

LH acknowledges the support from the Chinese National
Fundamental Research Program 2011CB921200 and
National Natural Science Funds for Distinguished Young Scholars.
XW acknowledges the financial support of start-up fund of Temple University and
computational support by the National Science Foundation through
TeraGrid resources provided by NICS under grant number [TG-DMR120045].

$^*$ To whom correspondence should be addressed: helx@ustc.edu.cn,xifanwu@temple.edu.


\begin{references}
\bibitem{HSE_PBE_U}
In the calculation based on PBEsol functional, an on-site Coulomb interaction
term $U=3.0$ eV and exchange interaction $J=1$ eV are used for technetium.
In our HSE hybrid functional calculations, a mixing parameter of $a$ = 0.3 is
adopted. A 500 eV planewave energy cutoff and $6\times6\times4$ $k$ mesh points converge
very well the results.
Ionic coordinates are considered to be fully relaxed until the Hellmann-Feynman
force is less then 1 meV/$\AA$.
\end{references}
\end{document}